\documentstyle[rotate,psfig]{mn2e}
\newcommand{\bugref}{\bibitem[\protect\citename{dummy }1893]{dum}}

\begin{document}

\title[Helical Fields in the Jets of BL~Lac Objects]
{Helical Magnetic Fields Associated with the Relativistic
Jets of Four BL~Lac Objects}

\author[Denise~C.~Gabuzda, \'Eamonn Murray \& Patrick Cronin]
{Denise~C.~Gabuzda, \'Eamonn Murray \& Patrick Cronin$^{1}$\\
$^1$Department of Physics, University College Cork, Cork, Ireland\\
}

\maketitle
\begin{abstract}
Evidence has been mounting that many of the transverse jet {\bf B} fields
observed in BL~Lac objects on parsec scales represent the dominant
vvroidal compoent of the intrinsic jet {\bf B} fields. Such fields could 
come about, for example, as a result of the  ``winding up'' of an initial 
``seed'' field with a significant longitudinal component by the rotation 
of the central accreting object. If this is the case, this should give 
rise to gradients in the rotation measure (RM) across the jets, due to 
the systematic change in the line-of-sight component of the jet {\bf B} 
field.  We present evidence for transverse RM gradients in four BL~Lac 
objects, strengthening arguments that the jets of these objects do indeed 
have toroidal or helical {\bf B} fields.  This underlines the view of the 
jets as fundamentally electromagnetic structures, and suggests that they
may well carry non-zero currents. It also provides a natural means to 
collimate the jets.

\end{abstract}

\section{Introduction}

BL~Lac objects are highly variable, appreciably polarized Active Galactic
Nuclei (AGN) that are observationally similar to radio-loud quasars in many
respects, but display systematically weaker optical line emission, whose
origin is not clear.  Very Long Baseline Interferometry (VLBI) polarization
observations of radio-loud BL~Lac objects have shown a tendency for
the polarization {\bf E} vectors in the parsec-scale jets to
be aligned with the local jet direction; since the jet emission is optically
thin, this implies that the corresponding magnetic {\bf B} field is transverse
to the jet (Gabuzda, Pushkarev, \&
Cawthorne 2000 \& references therein). This has usually been interpreted as
evidence for relativistic shocks that enhance the {\bf B}-field component
in the plane of compression, perpendicular to the direction of propagation
of the shock (Laing 1980; Hughes, Aller, \& Aller 1989).

However, as the quality of VLBI polarization images has improved and images
have become available for a larger number of objects at a larger number
of wavelengths, evidence has been building that many of the observed
jet {\bf B} fields actually correspond to the ``intrinsic'' {\bf B}
fields of the VLBI jets themselves.  For example,
in a number of BL~Lac objects, the observed
polarization vectors remain aligned with the jet over extensive
sections of the jet, even in the presence of appreciable bending
(0954+658: Gabuzda \& Cawthorne 1996; 1803+784: Gabuzda 1999; 0823+033 and
1749+701: Gabuzda \& Pushkarev 2002). A interpretation in the
framework of shock models requires the presence of a
series of relativistic shocks along the jet, each of which enhances
the local transverse {\bf B} field. Although theoretically possible,
this picture seems contrived.  An alternative is that the observed
transverse {\bf B} fields in these sources are associated with the
dominant toroidal component of the intrinsic jet {\bf B} fields, in
which case maintenance of the transverse field orientation as the jets
curve would be natural.

In addition, we might expect energetic shocks to be relatively compact
features, while some individual VLBI features that display transverse
{\bf B}-field structures are far from compact (e.g. 1219+285: Gabuzda et
al. 1994, 1803+784: Gabuzda \& Chernetskii 2003).  A number of jets with
alternating aligned
and orthogonal polarization vectors (orthogonal and aligned {\bf B}-field
structures) have also recently been discovered (e.g. OJ287: Gabuzda \&
G\'omez 2001; 1418+546: Gabuzda 2003).  In the
standard picture, these would correspond to fields ordered by
local phenomena at various places in the jet; however, a view in which
these alternating {\bf B} fields represent oscillations or instabilities
of a global jet {\bf B} field might be simpler.

Further, if a jet has a helical {\bf B} field, the observed {\bf B}-field
structure will depend on both the degree of dominance of the toroidal 
component and the viewing angle.  The viewing angle will play a role both 
geometrically and in terms of relativistic aberration effects.  For example, 
if the inferred observed {\bf B} field is everywhere transverse, with 
little or no sign of a longitudinal component near the edges, this would
require that we are viewing the jet nearly ``side-on'' (i.e., at 
$90^{\circ}$ to its axis) in its rest frame; if such a jet is viewed
at an angle that differs from $90^{\circ}$ to its axis in its rest frame,
we should see a central region of transverse {\bf B} field with regions
of longitudinal field appearing near the edges of the jet.
In fact, a number of compact AGN displaying just such ``spine+sheath'' 
{\bf B}-field structures have been observed.  The first example was 
1055+018 (Attridge, Roberts \& Wardle 1999), and such structures have 
since been observed in 0820+225 (Gabuzda, Pushkarev \& Garnich 2001), 
0745+241 (Pushkarev \& Gabuzda 2002), and 1652+398 (Gabuzda 2003c). 
This has sometimes been interpreted in the shock paradigm
as representing jets with series of shocks that are also interacting with a
surrounding medium (Attridge, Roberts \& Wardle 1999). 
However, as is indicated above, the hypothesis that some AGN jets 
have helical {\bf B} fields can also provide a simple explanation for 
these ``spine+sheath'' structures. 

Jet fields with an observed dominant toroidal component could come about
very naturally as a result of the ``winding up'' of an initial ``seed''
field with a significant longitudinal component by the rotation of the
central accreting
object (e.g.  Nakamura, Uchida, \& Hirose 2001; Lovelace et al. 2002;
Hujeirat et al. 2003; Lynden-Bell 2003; Tsinganos \& Bogovalov 2002).
Another intriguing possiblity is that the observed toroidal field components
develop in association with currents flowing in the jet (Pariev et al.
2003; Lyutikov 2003).  Indeed, if there is a dominant toroidal
{\bf B}-field component, basic physics tells us that in some sense there
{\em must} be currents flowing in the region enclosed by the field.

It is therefore of interest to identify robust observational tests that
can distinguish between transverse {\bf B} fields due to a toroidal
field component and due to shock compression. One such diagnostic
is provided by Faraday rotation -- rotation of the
observed plane of linear polarization due to propagation of the polarized
radiation through a magnetized plasma.  The rotation of the polarization
angle $\chi$ is determined by the observing wavelength $\lambda$, and
the density of free electrons $N(s)$ and the line-of-sight component of
the {\bf B} field in the plasma (e.g. Burn 1966):

\vspace*{-0.2cm}
\begin{eqnarray*}
\chi & = & \chi_0 +
 \frac{e^3\lambda^2}{8\pi m_e^2\epsilon_0c}\int N(s)\vec{B}(s)\cdot\,ds\\
 \chi & = & \chi_0 + \textrm{RM}\,\lambda^2
 \end{eqnarray*}

\noindent
where $\chi_0$ is the intrinsic polarization angle, $e$ is the electron
charge, $m_e$ is the electron mass, and the integral is taken over the line
of sight from the source to the observer.  Faraday rotation can be
identified via the characteristic quadratic dependence of the rotation of
$\chi$ on the observing wavelength; the coefficient of $\lambda^2$ is called
the rotation measure (RM).  In the case of a toroidal or helical jet {\bf B} 
field, we should observe a gradient in the observed Faraday rotation 
{\em across} the jet, due to the systematic change in the line-of-sight 
component of the {\bf B} field across the jet (see, e.g., Blandford 1993). 
Asada et al. (2002) claim to have detected such a gradient across the VLBI 
jet of 3C273, and interpreted this as evidence that this jet has a 
helical {\bf B} field.

In light of this result, we conducted a search for transverse RM gradients
among sources from a complete sample of northern BL~Lac objects
(K\"uhr \& Schmidt 1990) for which we had simultaneous VLBI polarization
data at 2, 4, and 6~cm obtained on the Very Long Baseline Array (VLBA).
We present here results for four of these sources in which we have
detected clear RM gradients transverse to the VLBI jets.  It is natural to 
interpret the observed RM gradients as revealing a toroidal or helical 
{\bf B} field associated with these VLBI jets, particularly given the other 
evidence that points in this direction.

\begin{figure*}
\hspace*{-1.0cm}
\mbox{
\rotate[r]{\psfig{file=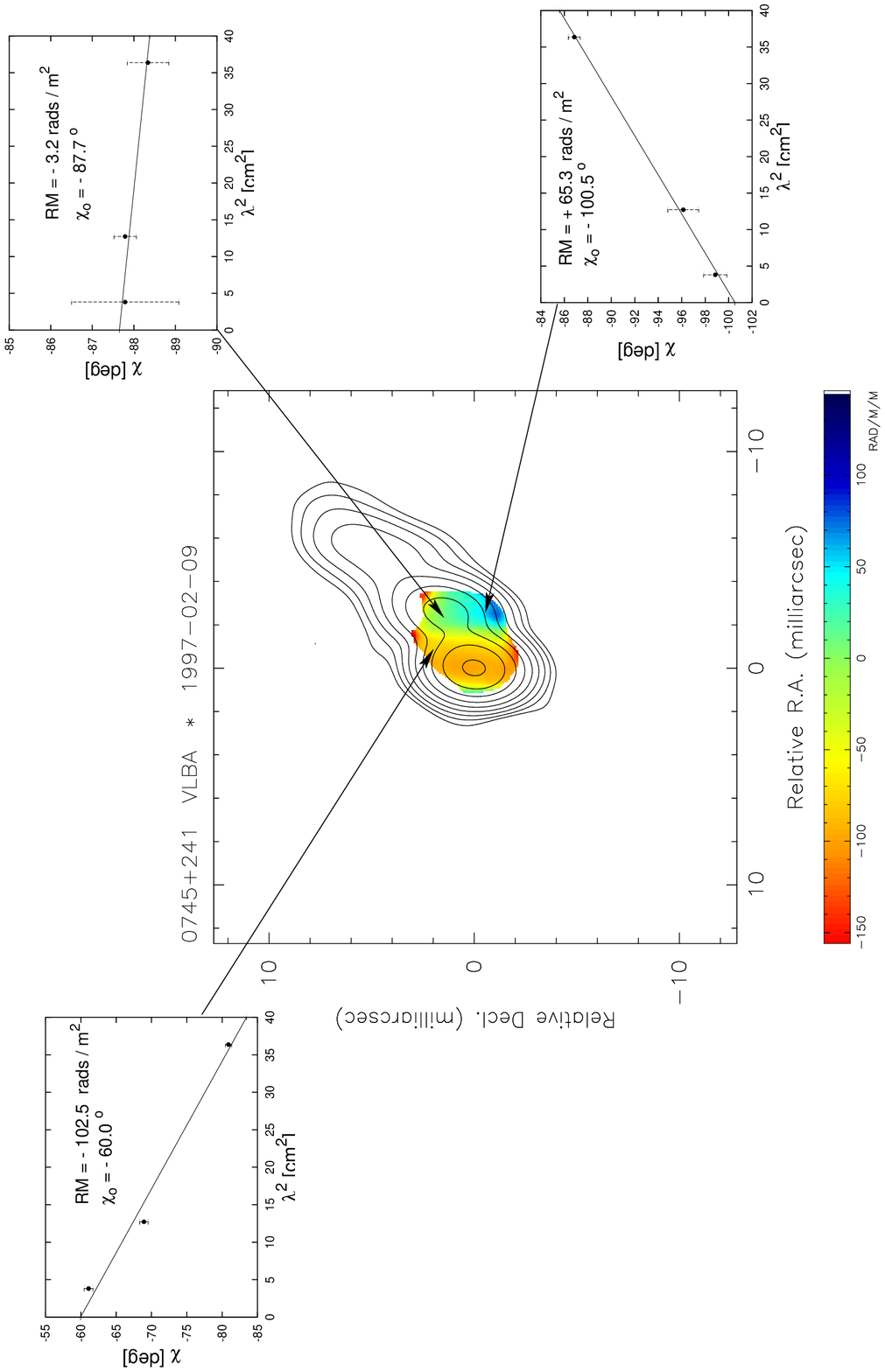,height=9.0cm}}
\rotate[r]{\psfig{file=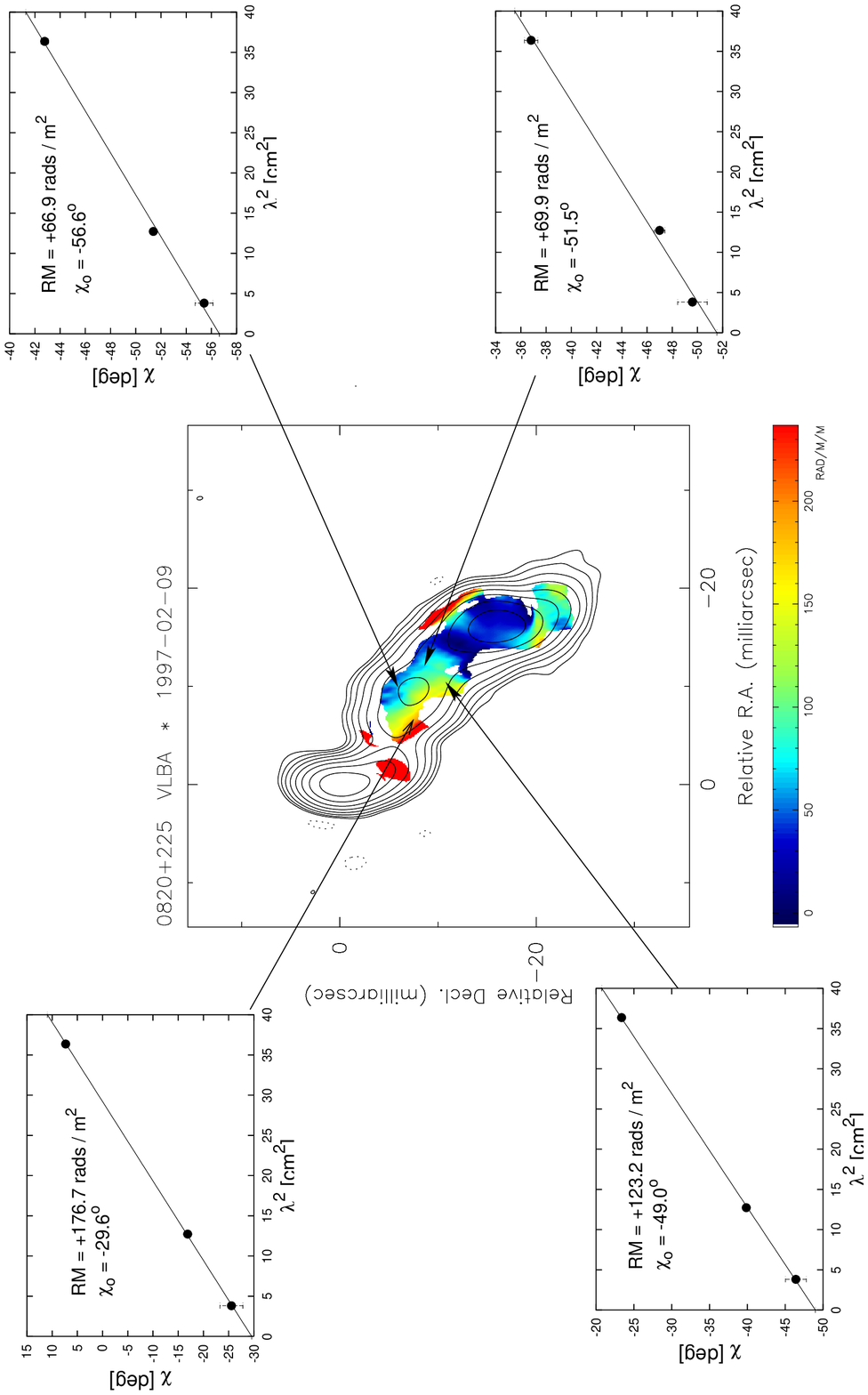,height=9.5cm}}
}
\hspace*{-1.0cm}
\mbox{
\rotate[r]{\psfig{file=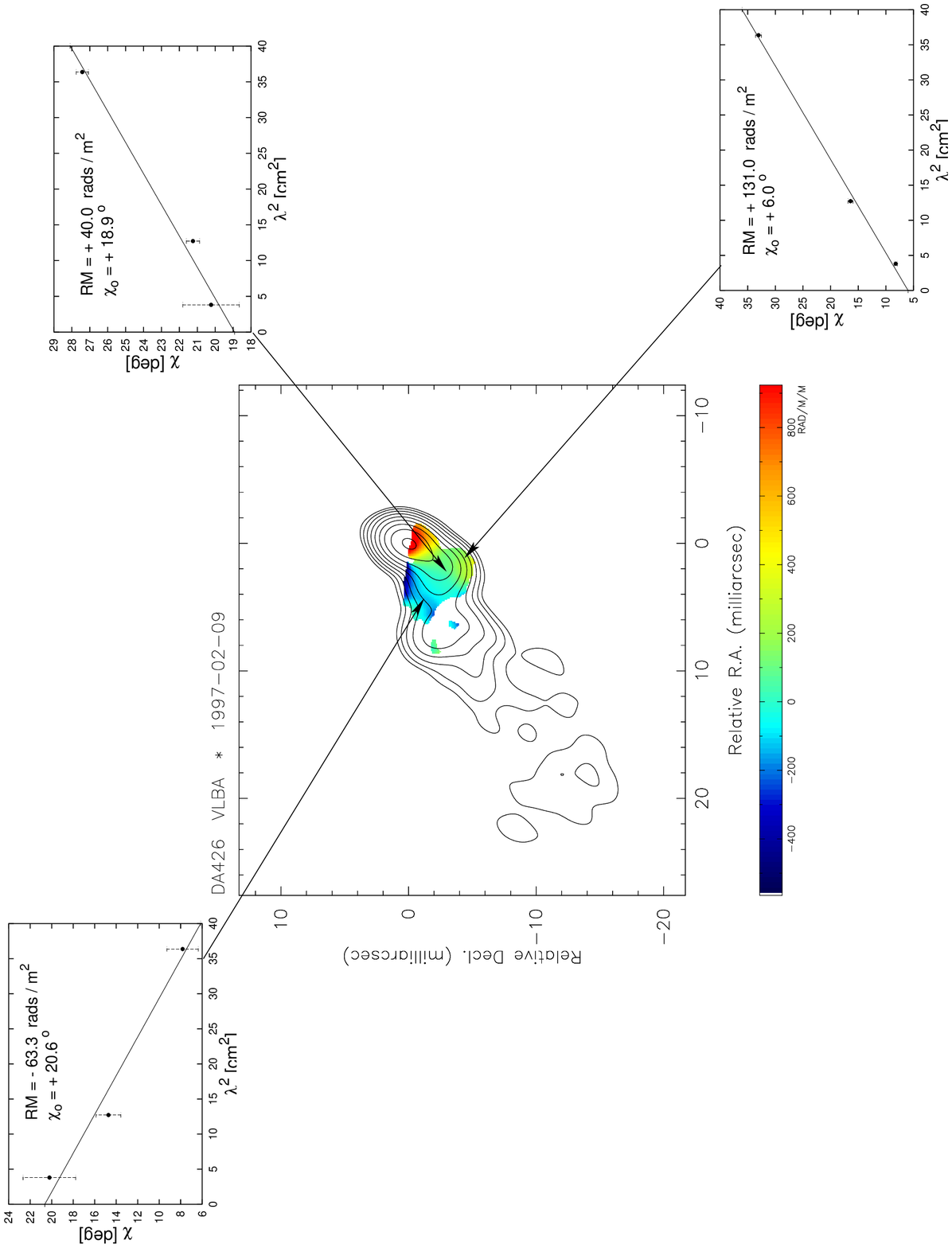,height=9.5cm}}
\rotate[r]{\psfig{file=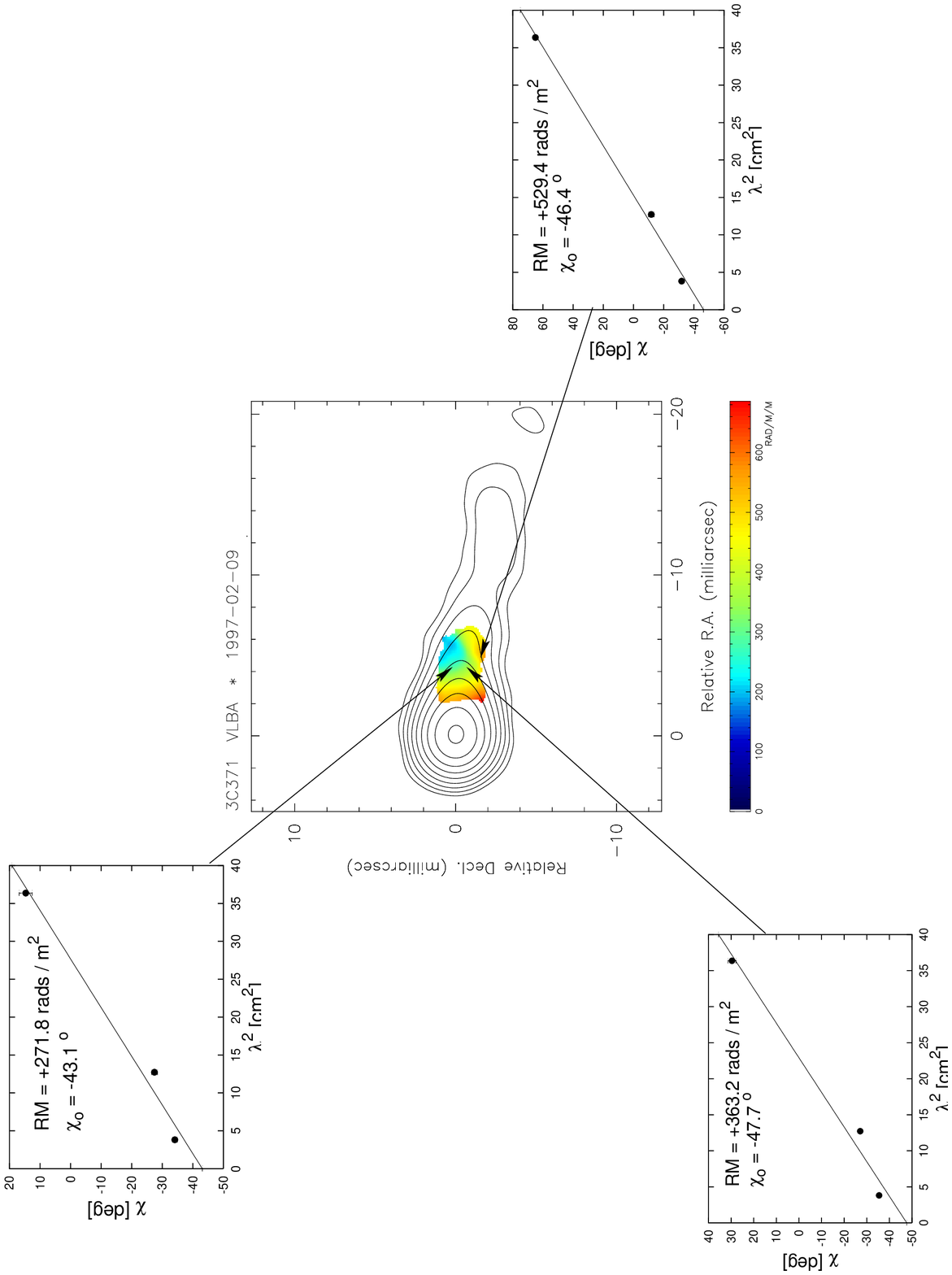,height=9.5cm}}
}
\caption{\label{fig:rmgrad} 6~cm VLBI $I$ maps of 0745+241 (top left),
0820+225 (top right), 1652+398 (bottom left) and 3C371 (bottom right),
with the parsec-scale RM distributions superimposed.
The accompanying graphs show plots of $\chi$ vs. $\lambda^2$ for several
locations across the VLBI jets. The $\chi$ errors shown are $1\sigma$. 
The convolving beams used for both the total-intensity and RM distributions
are $2.79\times 1.72$~mas in $PA = -4^{\circ}$ for 0745+241, 
$4.38\times 2.58$~mas in $PA=-6^{\circ}$ for 0820+225, $2.55\times 1.89$~mas
in $PA=-24^{\circ}$ for 1652+398 and $2.47\times 2.9$~mas
in $PA=-83^{\circ}$ for 3C371.}
\end{figure*}

\section{Observations and Reduction}

The polarization observations considered here were carried out on
February 9, 1997 at 6, 4 \& 2~cm using the NRAO Very Long
Baseline Array, as part of an ongoing study of the complete sample of
northern BL~Lac objects defined by K\"uhr \& Schmidt (1990). 
In all cases, the sources were observed in a ``snapshot'' mode,
with roughly 8--10 scans of each object spread out over the time the
source was visible with most or all of the VLBA antennas; the resulting
coverage of the $u$--$v$ plane was quite uniform. The data
reduction and imaging were done in the AIPS package using standard
techniques.

\begin{table}
\centering
\def\baselinestretch{1}
\caption[xx]{{\sc $Q$ and $U$ Noise Levels and $\chi$ Uncertainties}}
\label{tab:noise}
\begin{tabular}{ccccccc} \hline \hline
\multicolumn{2}{c}{2~cm} & \multicolumn{2}{c}{4~cm} &\multicolumn{2}{c}{6~cm}&\\
Q & U & Q & U & Q & U & Max $\Delta\chi$ \\
(mJy/ & (mJy/ & (mJy/ & (mJy/ & (mJy/& (mJy/ & (deg) \\ 
\phantom{x}beam) & \phantom{x}beam) & \phantom{x}beam) & \phantom{x}beam) & \phantom{x}beam)& \phantom{x}beam) &       \\ \hline
\multicolumn{7}{c}{0745+241}\\
0.40 & 0.48 & 0.23 & 0.26 & 0.17 & 0.23 & 6 \\
\multicolumn{7}{c}{0820+225}\\
0.21 & 0.21 & 0.20 & 0.13 & 0.18 & 0.17 & 5 \\
\multicolumn{7}{c}{1652+398}\\
0.45 & 0.47 & 0.19 & 0.18 & 0.22 & 0.23 & 8 \\
\multicolumn{7}{c}{3C371}\\
0.56 & 0.62 & 0.25 & 0.32 & 0.29 & 0.31 & 7 \\
\end{tabular}
\end{table}

The instrumental polarizations (``D-terms'') were derived from observations
of the unpolarized source 3C\,84 using the AIPS task LPCAL. 
The calibration of the absolute electric vector position angles (EVPAs) 
was determined using integrated polarization measurements obtained during 
a VLA session of about two hours duration on February 12, 1997. The primary 
EVPA calibrator used to calibrate the VLA polarization position angles
was 3C286 (assuming $2\chi = 66^{\circ}$, in the standard fashion).  The 
VLBI EVPAs were calibrated by requiring that the EVPA for the total VLBI 
polarization of a compact source coincide with the EVPA for the VLA core.  
The source used for this purpose were 1823+568, for which more than 90\% of the 
VLA core polarization is present on VLBI scales (e.g. Gabuzda et al. 1992). 
The near 
simultaneity of the VLA and VLBA observations minimizes the possibility 
of variations in the source polarization between the dates of the VLA and 
VLBA observations. We estimate that these EVPA calibrations are accurate to 
within about $3^{\circ}$.

Note that any errors in the EVPA calibration affect all polarization
angles in all regions of all the sources in the same way, and so could not
lead to the presence of spurious Faraday-rotation gradients in the source
structure.

We made maps of the distribution of the 2, 4 and 6-cm total intensity $I$ 
and Stokes parameters $Q$ and $U$ with matched resolutions corresponding 
to the 6-cm beam; in the case of 0820+225, the beam used for the 
matched-resolution maps was 1.5 times the size of the 6-cm beam. The 
$Q$ and $U$ maps were than used to construct the distributions of the 
polarized flux ($p = \sqrt{Q^2 + U^2}$) and polarization angle 
($\chi = \frac{1}{2}\arctan \frac{U}{Q}$), as well as accompanying ``noise 
maps,'' using the AIPS task COMB. The formal uncertainties written in the 
output $\chi$ noise maps were calculated in COMB using the rms noise levels on 
the input $Q$ and $U$ maps, summarized in the Table.

Before making the final RM maps using the AIPS task RM, we first removed the
contribution of the known integrated (predominantly Galactic, i.e., foreground; 
Pushkarev 2001) Faraday rotation at each wavelength, so that any remaining 
non-zero rotation measure should be due to thermal plasma in the vicinity 
of the AGN. The task RM has the option of blanking output values in pixels
for which the $\chi$ errors at any of the three wavelengths used are larger 
than a specified value. The maximum $\chi$ uncertainties for a single 
pixel applied to construct the rotation-measure maps shown in Fig.~1 are 
also indicated in the Table.

For the purposes of determining the $\chi$ values for individual regions 
in the VLBI jet, we found the mean within a $3\times 3$~pixel 
($0.3\times 0.3$~mas) area (i.e., nine elements) at the corresponding 
location in the map; the errors in these polarization angles were taken 
to be the rms deviation from this mean value. 

\section{Results}

Fig.~1 shows the 6-cm VLBI $I$ images of four objects displaying
transverse RM gradients with color images of their parsec-scale RM
distributions superposed. The convolving beams used in each case are
indicated in the figure caption. The RM gradients across the VLBI jets are
visible by eye. The accompanying plots show the observed polarization
position angles, $\chi$, as a function of the observing wavelength squared,
$\lambda^2$, for individual locations in the VLBI jets. The $\chi$ errors
shown are $1\sigma$. In all cases, the observed $\chi$ values are in good
agreement with the $\lambda^2$ law expected for Faraday rotation.

\begin{figure*}
\hspace*{-1.0cm}
\mbox{
\rotate[r]{\psfig{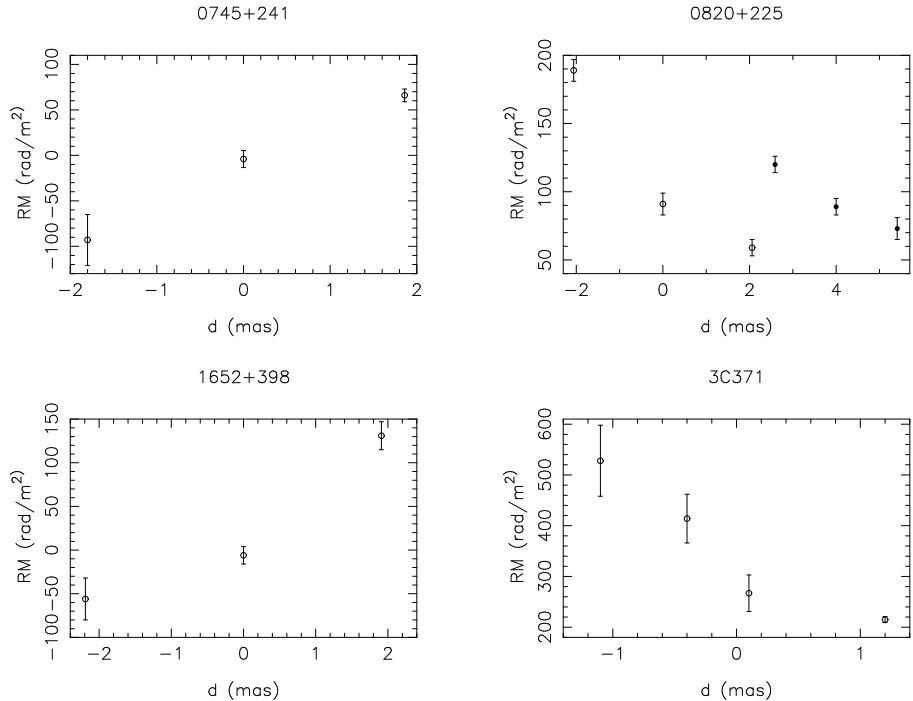}}
}
\caption{\label{fig:rmplots} Plots of observed rotation measure as a function
of transverse distance $d$ from a reference point $P$ near the central jet 
spine for cuts along the indicated position angles $PA$ across the jets of 
the four objects shown in Fig.~1. Top left (0745+241): $P = 
(-1.5~\textrm{mas}, 0.7)$~mas, $d$ measured along $PA= 35^{\circ}$, with 
northeast $d$ being negative. Top right (0820+225): open circles show the
results for $P = (-8.6~\textrm{mas}, -6.7)$~mas, $d$ measured along 
$PA=-60^{\circ}$, with southeast $d$ being negative; filled circles shown 
the results for $P = (-11.6~\textrm{mas}, -9.6)$~mas, $d$ measured along 
$PA=-45^{\circ}$, with southeast $d$ being negative. The distances for the
latter three RM values have been arbitrarily shifted 4.0~mas to ease viewing
of the results for both cuts in a single plot. Bottom left (1652+398): 
$P = (3.0~\textrm{mas}, -3.2)$~mas, $d$ measured along $PA=50^{\circ}$, with 
northeast $d$ being negative.  Bottom right (3C371): $P = (5.4~\textrm{mas}, 
-0.5)$~mas, $d$ measured along $PA=0^{\circ}$, with south $d$ being negative.
All errors are $2\sigma$.}
\end{figure*}

Fig.~2 shows plots of the observed RMs with their errors as a function 
of approximate transverse distance from the central spine of the VLBI jets of 
each of the four objects; the point from which the distances were measured
and the position angle along which they were measured are indicated in the
figure caption. The errors in the
RMs were estimated in two ways: (i) in the same way as for the $\chi$
values, by calculating the mean RM within the corresponding 
$3\times 3$~pixel ($0.3\times 0.3$~mas) area (i.e., nine elements) in the 
map and assigning the rms deviation about this mean as the RM error, and 
(ii) from the linear fit to $\chi$ vs. $\lambda^2$. 
In most cases, these two methods gave comparable errors, but if 
there was a significant difference in the two errors, we adopted 
the larger of the two as the error in the RM, in order to help ensure that
the errors were not severely underestimated. Fig.~2 confirms that the 
transverse RM gradients are highly significant. The 
differences between the two RM values on either side of the central spine 
of the jets are $8.5\sigma$ for 0745+241, $18.6\sigma$ and $6.7\sigma$ 
for 0820+225 (for transverse cuts roughly 10~mas and 15~mas from the core, 
respectively), $9.4\sigma$ for 1652+398 and $8.2\sigma$ for 3C371. This 
leaves no doubt of the reality of the observed transverse RM gradients.

\section{Discussion}

In order to obtain the RM distributions shown, it was necessary to
convolve the 2-cm $Q$ and $U$ images with a beam that was roughly
three times the beam corresponding to the intrinsic resolution of the
2-cm data. This might give rise to concerns that a blending of 2-cm and
4-cm polarized components located within the 6-cm beam could lead to
the appearance of false structure in the resulting RM maps. In none of
the four sources considered here was there transverse structure
in the jet polarization in the regions shown in the RM maps that was 
evident at the shorter wavelengths, but not at 6~cm. Two additional
facts convince us that the observed RM gradients are not spurious
structures resulting from the need to use a single convolving beam when
constructing the images at all three wavelengths: (i) the gradients 
are systematic and monotonic, and (ii) in all cases, the
observed $\chi$ values display a clear $\lambda^2$ dependence, 
demonstrating that we are seeing the manifestation of Faraday rotation.

An independent 2+4+6~cm VLBA RM map for 1652+398 (= Mrk501)
for epoch April 1998 analyzed by Croke et al. (2004) shows a
transverse RM gradient with the same sense and covering roughly the same
range of rotation measures as the RM map of this source shown in Fig.~1.
A preliminary analysis of 6, 13, and 18-cm VLBA polarization data for this
source for May 1998 also displays a transverse RM gradient in the same sense
(Croke et al., in preparation). These independent RM distributions serve to
confirm the reality of the gradients we have detected in this source.

The transverse rotation-measure gradients observed in 0745+241 and 
1652+398 (left panels in Fig.~1) are positive on one side of the jet spine 
and negative on the other. This behaviour is expected if
we are viewing the jet nearly orthogonal to its axis {\em in the source frame}.
If the Lorentz factor for the jet's motion is $\gamma$, photons emitted
at $90^{\circ}$ to the jet axis in the source frame will be observed in
the observer's frame to propagate at an angle $\theta = 1/\gamma$ to the
jet axis.  Since we believe that compact radio-loud AGN such as BL~Lac
objects represent precisely those sources whose jets are viewed at small
angles to the line of sight, it is quite natural to observe this behaviour
in the rotation-measure gradients. If the jet is viewed at an angle somewhat
different than $90^{\circ}$ to the jet axis in the source frame, this will
lead to an offset in the observed RM gradients, so that the observed RM
will not be symmetrical about a rotation measure of zero; this can
qualitatively explain the RM gradients observed for 0820+225 and 3C371 
(right panels in Fig.~1).

Because relativistic boosting of the jet emission changes the relative
strength of the {\bf B}-field components along and orthogonal to the
line of sight, the toroidal {\bf B}-field component will become dominant
in the observer's frame, even if the toroidal and poloidal {\bf B}-field
components in the source frame are comparable (e.g., Blandford \& K\"onigl
1979; this is discussed in more detail by Lyutikov, Pariev \& Gabuzda, in
preparation).  This could provide a natural explanation for why
Faraday-rotation gradients transverse to the jet axis appear to be fairly
common among the objects we have studied.

The presence of a helical {\bf B} field carried outward with the
relativistic outflows provides a natural mechanism for the collimation
of these outflows, which in some cases remain extremely well collimated
out to kiloparsec scales. Although it is difficult to prove conclusively,
the presence of toroidal or helical {\bf B} fields genetically related to
the relativistic jets suggests that a significant fraction of their energy 
may be carried by the ordered component of the magnetic field, rather than 
by the hydrodynamical energy of the flow.

Another intriguing possibility is that a connection will emerge between the
evidence we have found for helical {\bf B} fields associated with these
four BL~Lac objects and the recent detection of circular polarization in 
the core regions of a number of AGN (Homan, Attridge \& Wardle 2001, Gabuzda \&
Vitrishchak 2004).  It is possible that both the helical fields and the
circular polarization are associated with the rotation of the central
black hole and accretion disk; in this case, there should be a direct
relationship between the transverse RM gradients (which reflect the
direction of the rotation) and the sign of the circular polarization. 
There are two sources for which both transverse RM gradients and 
parsec-scale circular polarization have been detected: 3C273 and 1807+398 (= 
3C371). In the case of 3C273, the jet propagates to the southwest, and the 
RM is more positive along the northwestern edge of the jet, implying that the 
rotation of the accretion disk is {\em clockwise} as viewed by the observer 
(Asada et al. 2002); the VLBI core circular polarization measured by 
Homan \& Wardle (1999) was {\em negative}.  In the case of 3C371, the VLBI jet 
propogates toward the west and the RM is more positive along the southern 
edge of the jet, implying that the rotation of the accretion disk is 
{\em counterclockwise} as viewed by the observer; the VLBI core circular 
polarization measured by Gabuzda \& Vitrishchak (2004) was {\em positive}.  
Thus, these two cases are consistent with both the transverse RM gradients
and the parsec-scale circular polarization having a common origin, but it 
is obviously impossible to know if this is a coincidence on the basis of only
these two sources. It will be of considerable interest to carry out joint 
analyses of the RM gradients and core circular polarizations for additional 
sources as more data become available to see if these two quantities are, 
indeed, related.

\section{Conclusion}

Our analysis has demonstrated the presence of Faraday-rotation gradients 
transverse to the VLBI jets of four BL~Lac objects.  These transverse RM 
gradients provide compelling new evidence that the jet {\bf B} fields of 
these objects are dominated by the toroidal component, indicating that 
the {\bf B} fields themselves are either toroidal or helical. This 
represents the first direct observational evidence for toroidal or helical
{\bf B}-field structures associated with the jets of AGN.

This has fundamental implications for our understanding of the launching 
and propagation of these jets; in particular, it is clear that we must 
think of the jets as fundamentally electromagnetic structures.  These 
results provide support for models in which the launching of the jets is 
associated with a seed {\bf B} field initially threading the accretion disk 
that is wound up by the rotation of the disk and central black hole. This 
has seemed like an obvious possibility from the theoretical point of view, 
but until now, there has been no observational evidence that directly 
supported such models.  In addition, the presence of helical fields would 
provide a natural mechanism for collimating the jets.

An analysis of the parsec-scale RM distributions for remaining sources
in the K\"uhr \& Schmidt (1990) sample of BL~Lac objects is underway, and 
results will be presented in a future paper, where we will also consider 
the question of to what extent transverse RM gradients are characteristic 
of the sample as a whole.  We are also in the process of obtaining new 
multi-wavelength VLBA polarization observations for all the sources in 
this sample of BL~Lac objects, designed to be better optimized for 
Faraday-rotation studies; these new observations should enable us to both 
confirm the results obtained thus far and to study the parsec-scale RM 
distributions with appreciably better resolution.

\section{Acknowledgements}

We thank the referee for useful and pertinent comments, as well as for 
prompt reviewing of the manuscript.
The National Radio Astronomy Observatory is operated by Associated
Universities, Inc., under cooperative agreement with the NSF.

\clearpage

\end{document}